\documentclass[superscriptaddress,pre,twocolumn]{revtex4-2}
\usepackage{amsmath}
\usepackage{amssymb}
\usepackage{amscd}
\usepackage{graphicx}
\usepackage{bbm}
\usepackage{dsfont}
\usepackage{datetime}
\usepackage{color}
\usepackage[normalem]{ulem}

\longdate

\usepackage{tikz}
\usepackage{upgreek}

\begin{document}
\title{Spontaneous Flows and Quantum Analogies in Heterogeneous Active Nematic Films
}
\author{Alexander J.H. Houston}
\email{Corresponding author:Alexander.Houston@glasgow.ac.uk}
\author{Nigel J. Mottram}
\email{Nigel.Mottram@glasgow.ac.uk}
\affiliation{School of Mathematics and Statistics, University Place, Glasgow, G12 8QQ, United Kingdom}

\begin{abstract}
Incorporating the inherent heterogeneity of living systems into models of active nematics is essential to provide a more realistic description of biological processes such as bacterial growth, cell dynamics and tissue development. Spontaneous flow of a confined active nematic is a fundamental feature of these systems, in which the role of heterogeneity has not yet been considered. We therefore determine the form of spontaneous flow transition for an active nematic film with heterogeneous activity, identifying a correspondence between the unstable director modes and solutions to Schr\"{o}dinger's equation. We consider both activity gradients and steps between regions of distinct activity, finding that such variations can change the signature properties of the flow. The threshold activity required for the transition can be raised or lowered, the fluid flux can be reduced or reversed and interfaces in activity induce shear flows. In a biological context fluid flux influences the spread of nutrients while shear flows affect the behaviour of rheotactic microswimmers and can cause the deformation of biofilms.
All the effects we identify are found to be strongly dependent on not simply the types of activity present in the film but also on how they are distributed.
\end{abstract}
\maketitle

\section{Introduction}
\label{sec:intro}
Collective motion is a ubiquitous feature of living systems. It can be observed on many scales, from flocks of birds \cite{bialek2012statistical} and schools of fish \cite{lopez2012behavioural} to swarms of bacteria \cite{wensink2012meso}. This last instance is an example of an active nematic; biological or synthetic materials composed of orientationally ordered motile constituents that drive the system away from equilibrium \cite{ramaswamy2010mechanics,marchetti2013hydrodynamics,doostmohammadi2018active} and which may be modelled by the addition of an active stress to the hydrodynamics of passive nematics \cite{de1993physics}. Further examples include suspensions of synthetic microtubules \cite{sanchez2012spontaneous}, bacteria in liquid crystalline environments \cite{zhou2014living}, tissues \cite{saw2017topological} and cell monolayers \cite{duclos2017topological}. In this context collective motion takes the form of coherent flows on scales larger than an individual and such flows have implications for organ development \cite{mclennan2012multiscale}, wound healing \cite{poujade2007collective} and the formation  \cite{stoodley1994liquid,kjelleberg2002there,purevdorj2002influence} and spread of nutrients within biofilms \cite{hall2004bacterial}. They also result in the spontaneous motion \cite{yao2022topological,loewe2022passive,houston2023active} and rotation \cite{ray2023rectified,houston2023colloids} of colloids and the self-propulsive \cite{narayan2007long,giomi2013defect,giomi2014defect,binysh2020three} and self-orienting \cite{houston2022defect} dynamics of topological defects. The latter have been shown to be pivotal to biological functionality in epithelial cell apoptosis \cite{saw2017topological}, the formation of bacterial colonies \cite{doostmohammadi2016defect,dell2018growing,basaran2022large} and biofilms \cite{yaman2019emergence} and morphogenesis \cite{maroudas2021topological,guillamat2022integer}.

The ability of a confined active nematic to, above a threshold level of activity, spontaneously transition from an aligned quiescent state to a distorted flowing one, is a key hallmark of its non-equilibrium nature, without parallel in passive systems. Although it bears some resemblence to the Freedericksz transition \cite{de1993physics}, it stems from the fundamental hydrodynamic instability of active nematics \cite{aditi2002hydrodynamic,ramaswamy2010mechanics} due to their active stresses. This flow transition was theoretically predicted in 2005 \cite{voituriez2005spontaneous} and has since been confirmed numerically \cite{marenduzzo2007steady} and experimentally \cite{duclos2018spontaneous}. In its wake there have been numerous studies of the role confinement can play in modifying the dynamics of active nematics \cite{shendruk2017dancing,duclos2017topological,opathalage2019self}.

However, active biological systems are not homogeneous, nor do they exist in isolation, but rather in a complex environment. Active non-uniformities can arise due to the presence of distinct bacterial phenotypes \cite{meacock2021bacteria} or differing cell-cell interactions \cite{balasubramaniam2021investigating}, while interfaces with a passive environment appear in bacterial invasion \cite{meacock2021bacteria,burmolle2006enhanced,nadell2015extracellular}, wound closing \cite{brugues2014forces} and the motion of cells through extracellular matrices \cite{friedl2009collective,friedl2012classifying}. Furthermore, the generation of light-activated materials \cite{nakamura2014remote,ross2019controlling,ruijgrok2021optical} has enabled spatio-temporal control of active nematics by using variations in light intensity to pattern the activity \cite{zhang2021spatiotemporal,shankar2024design}. Both these in vivo and in vitro motivations call for the incorporation of heterogenous activity in active nematics and an understanding of its impact.

Accordingly, we consider here spontaneous flow transitions in heterogenous active nematic films, using a correspondence with Schr\"{o}dinger's equation. Following this correspondence, we consider the effect of activity gradients, steps, wells and barriers. We find that the presence of heterogeneity and interfaces in active systems can lower the threshold to flow transitions and result in complex flows, with multiple changes in flow direction. Crucially, the nature of the transition is highly sensitive not just to the type of activity present but also where it is located.

\section{Methods}

The starting point for our analysis is the Ericksen-Leslie equations for an active nematic in terms of its director field $\mathbf{n}$ and fluid flow field $\mathbf{u}$,
\begin{align}
    \nabla\cdot\mathbf{u}&=0, \label{eq:Incompressable}\\
    -\nabla p+\mu\nabla^2\mathbf{u}+\nabla\cdot\mathbf{\sigma}&=0,
    \label{eq:Stokes}\\
    \partial_t\mathbf{n}+\mathbf{u}\cdot\nabla\mathbf{n}+\mathbf{\Omega}\cdot\mathbf{n}&=\frac{1}{\gamma}\mathbf{h}-\nu\left[\mathbf{D}\cdot\mathbf{n}-(\mathbf{n}\cdot\mathbf{D}\cdot\mathbf{n})\mathbf{n}\right],
    \label{eq:NematoHydro}
\end{align}
where $p$ is the pressure, $\mu$ the Newtonian viscosity, $\sigma$ the total stress, $\gamma$ the rotational viscosity, $h$ the molecular field,  $\nu$ the flow alignment parameter and $D_{ij}=\frac{1}{2}\left(\partial_iu_j+\partial_ju_i\right)$ and $\Omega_{ij}=\frac{1}{2}\left(\partial_iu_j-\partial_ju_i\right)$ are the symmetric and antisymmetric parts of the flow gradients respectively. In the present work, we consider flow-aligning nematics, for which $\nu<-1$. We employ a one elastic constant approximation in which the Frank free energy density has the form $F=\frac{K}{2}\partial_in_j\partial_in_j$, from which $\mathbf{h}$ is found to be
\begin{align}
    h_i&=-\frac{\delta F}{\delta n_i}+\frac{\delta F}{\delta n_j}n_jn_i\nonumber\\
    &=K\left[\nabla^2n_i-n_in_j\nabla^2n_j\right].
\end{align}
The total stress is the sum of elastic and active stresses, given by
\begin{align}
    \sigma_{ij}=&\frac{1}{2}\left(n_ih_j-h_in_j\right)+\frac{\nu}{2}\left(n_ih_j+h_in_j\right)\nonumber\\
    &\qquad-K\partial_in_k\partial_jn_k-\zeta n_in_j,\label{stress}
\end{align}
where the final term is the active stress and the remaining terms constitute the elastic contribution. The active contribution to the stress tensor is weighted by the activity $\zeta$, a phenomenological parameter which is positive for extensile active nematics and negative for contractile active nematics.

\begin{figure}
    \centering
    \begin{tikzpicture}[scale=1.6,>=stealth]
        \node[anchor=south west,inner sep=0] at (0,0)
{\includegraphics[width=0.95\linewidth, trim = 0 0 0 0, clip, angle = 0, origin = c]{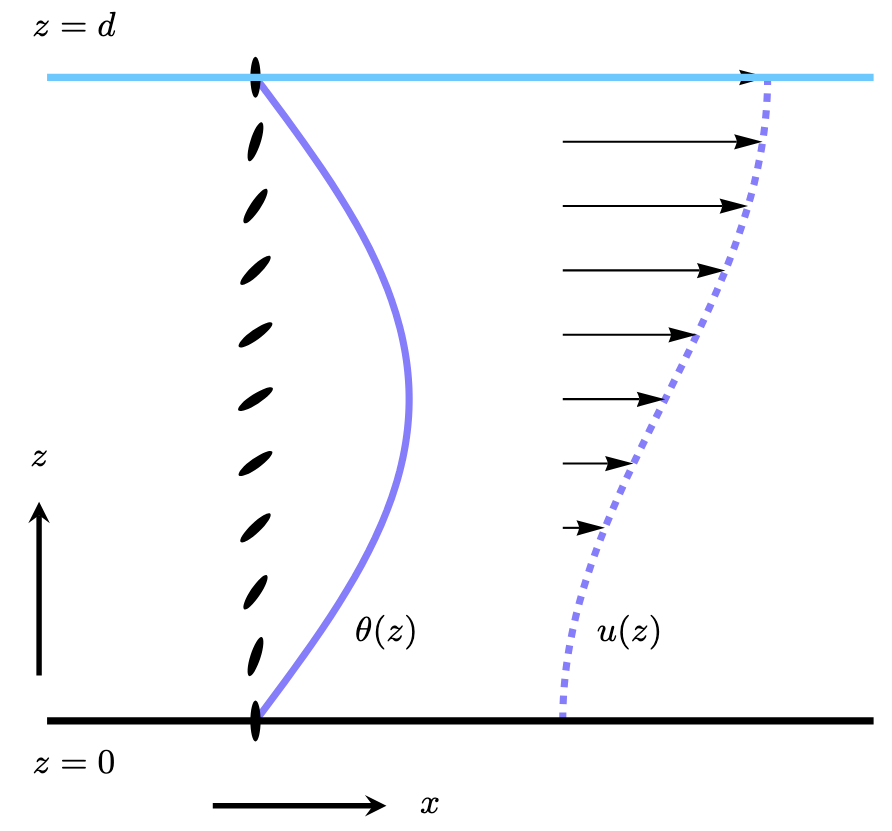}};
    \end{tikzpicture}
    \caption[The \textcolor{red}{unstable mode associated with the} spontaneous flow transition for a homogeneous active nematic film.]{The unstable mode associated with the spontaneous flow transition for a homogeneous active nematic film. The film sits upon a substrate at $z=0$ and has a free surface at $z=d$. The director is shown along a line through the film as black ellipses, with the corresponding variation in director angle shown as a solid line. The flow velocity is indicated by the arrows and accompanying dashed line.
    }
    \label{fig:homogeneousInstability}
\end{figure}

We consider a film geometry, with a solid substrate at $z=0$ and a free surface at $z=d$, shown in Figure \ref{fig:homogeneousInstability}.  Following \cite{voituriez2005spontaneous}, we employ a quasi-one-dimensional description possessing translational invariance along the $x$- and $y$-directions and a director field that remains in the $xz$-plane, such that $\mathbf{n}=\sin\theta(z)\mathbf{e}_x+\cos\theta(z)\mathbf{e}_z$. Incompressibility enforces rectilinear flow, such that $\mathbf{u}=u(z)\mathbf{e}_x$. We consider homeotropic anchoring, corresponding to $\theta=0$, on both the substrate and the free surface, a no-slip condition for the flow at the substrate and a stress-free condition at the free surface. We take the anchoring strengths at the free surface and substrate to be equal and infinitely strong. Such an approximation is commonplace, and supported by large experimentally-measured free surface anchoring strengths \cite{chiarelli1983determination,slavinec1997determination}. Using these conditions and linearising about $\theta\equiv 0$ allows the director dynamics equation \eqref{eq:NematoHydro} to be reduced to
\begin{equation}
    \partial_t\theta=\frac{K}{\gamma}\partial^2_z\theta+\frac{1-\nu}{2}\partial_zu,
    \label{eq:DirectorDynamicsFlow}
\end{equation}
with the quiescent state becoming critically unstable to a distorted, flowing one when $\partial_t\theta=0$ \cite{voituriez2005spontaneous}.

To solve for the flow we begin by noting that the $z$-component of \eqref{eq:Stokes} integrates to give the pressure as
\begin{equation}
    p=\sigma_{zz}+p_0,
\end{equation}
where $\sigma_{ij}$ is the $ij$-component of the active-plus-elastic stress tensor and $p_0$ is a constant reference pressure. Taking the $x$-component of \eqref{eq:Stokes} yields
\begin{equation}
    \partial_z\left(\mu\partial_zu+\sigma_{xz}\right)=0.
    \label{eq:FlowStress}
\end{equation}
The parenthetical term is the total stress, which, since it is zero at the free surface, must vanish everywhere. It then follows from \eqref{eq:FlowStress} that the flow is given by
\begin{equation}
    u(z)=-\frac{1}{\mu}\int_0^z\sigma_{xz}\text{d}z',\label{usol}
\end{equation}
where here we have used the no-slip boundary condition at the substrate. This flow solution holds provided $\sigma_{xz}$ is continuous and so will need to be modified when we come to consider activity steps in the `Layered active systems' subsection of the Results.

From \eqref{stress}, and the  form of the director used in the present work, the relevant components of the stress tensor are
\begin{align}
    \sigma_{zz}&=-\left[K\left(\partial_z\theta\right)^2+K\nu\sin\theta\cos\theta\partial_z^2\theta+\zeta\cos^2\theta\right],\\
    \sigma_{xz}&=-\left[\frac{K}{2}(1-\nu\cos2\theta)\partial_z^2\theta+\zeta\sin\theta\cos\theta\right],
\end{align}
which linearise to
\begin{align}
    \sigma_{zz}&=-\zeta,\\
    \sigma_{xz}&=-\left[\frac{K(1-\nu)}{2}\partial_z^2\theta+\zeta\theta\right].
    \label{eq:SigmaXZLinearised}
\end{align}
Substituting the explicit form of $\sigma_{xz}$ given in \eqref{eq:SigmaXZLinearised} into \eqref{usol} and using this in \eqref{eq:DirectorDynamicsFlow}, we find that the condition for critical instability becomes
\begin{equation}
    \eta\partial^2_z\theta+\zeta(z)\theta=0,\qquad \eta=\frac{K\left[4\mu+\gamma(1-\nu)^2\right]}{2\gamma(1-\nu)}.
    \label{eq:SchrodingerDirector}
\end{equation}
That is, the linearly unstable director modes satisfy the time-independent Schr\"{o}dinger equation. The activity plays the role of a potential, with extensile ($\zeta>0$) and contractile ($\zeta<0$) regions corresponding to classically allowed and forbidden domains respectively. This correspondence is not universal but rather dictated by the boundary conditions. It appears as described here because for homeotropic anchoring the distortion is, to linear order, pure bend, which is unstable in extensile systems \cite{ramaswamy2010mechanics}. 
If planar anchoring were enforced, the results in this paper would still hold, although the instability would be for contractile systems, would occur through a splay distortion, and the correspondence with allowed and forbidden domains would be reversed. This reflects part of the symmetry which exists between regions of material parameter space for two-dimensional, although not three-dimensional, confined active nematics \cite{edwards2009spontaneous}.

Finally, we may use \eqref{eq:SchrodingerDirector} in combination with \eqref{eq:FlowStress} to express the fluid flow as
\begin{equation}
    u(z)=\frac{4}{4\mu+\gamma(1-\nu)^2}\int_0^z\zeta\theta\,\text{d}z'.
    \label{eq:FlowDirector}
\end{equation}
In the classical situation of uniform activity \cite{voituriez2005spontaneous}, solving \eqref{eq:SchrodingerDirector} gives a transition to a distorted state through a single sinusoidal mode and at a critical activity $\zeta_{\text{c}}$, given by
\begin{align}
    \theta(z)\sim\sin\left(\sqrt{\frac{\zeta_{\text{c}}}{\eta}}z\right), &\ \   u(z)\sim 1-\cos\left(\sqrt{\frac{\zeta_{\text{c}}}{\eta}}z\right)\\
    \zeta_{\text{c}}&=\frac{\uppi^2\eta}{d^2},
    \label{eq:homogeneousInstability}
\end{align}
as shown in Figure \ref{fig:homogeneousInstability}. Note that for a continuous activity profile the viscous, elastic and active stresses are all directly proportional to the director angle and accordingly vanish at both the free surface and the substrate. The exception to this is when discontinuous steps in the activity are present, since then, as we shall see, the viscous stress will not generically be zero at the substrate due to induced shear flows.

\section{Results}

\subsection{Constant activity gradient}
A quintessential form of active heterogeneity is a constant activity gradient, which captures the leading order effect of smooth variations in activity. Such gradients might arise due to structured variation in the active agents, such as bacteria forming into layers with younger bacteria on top \cite{you2019mono}, or in their concentration, as occurs for microswimmers in confinement \cite{spagnolie2012hydrodynamics}. Variable activity could also exist because of spatial variation in a requirement for active behaviour. For example, nutrients could be concentrated at the substrate or for light-activated systems \cite{zhang2021spatiotemporal} the light intensity could be highest at the free surface, decaying into the layer. Lastly, variations in activity could be present as a result of the active agents adaptively changing their activity in response to some varying environmental feature. While it is perhaps most natural to consider such smooth variations in the context of a single active species, they can also arise in multi-phase active systems, including extensile-contractile mixtures. We therefore allow the activity to be both positive and negative at different locations within the same system, while noting that a true multi-phase system will incorporate both smooth variations and the activity jumps that we will consider in the next section.

\begin{figure*}
    \centering
    \begin{tikzpicture}[scale=1.6,>=stealth]
        \node[anchor=south west,inner sep=0] at (0,0)
{\includegraphics[width=0.95\linewidth, trim = 0 0 10 0, clip, angle = 0, origin = c]{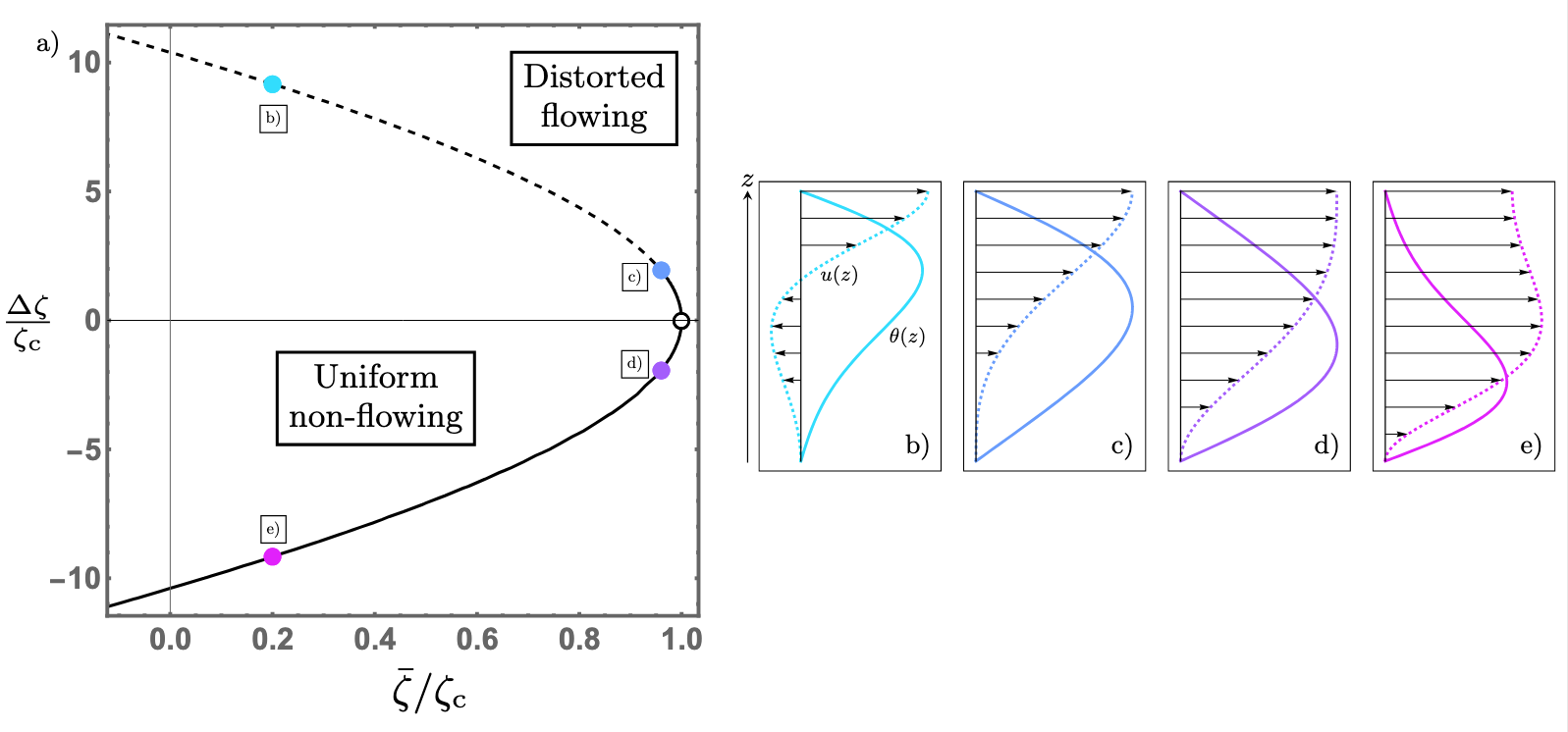}};
    \end{tikzpicture}
    \caption[The spontaneous flow transition of an active nematic film with an activity gradient.]{The spontaneous flow transition of an active nematic film with an activity gradient. The open circle corresponds to the critical activity in a homogeneous system.
    a) The line along which linear instability occurs for an activity profile given by Eq. \eqref{eq:LinearActivity}. The solid portion indicates the regime for which unidirectional flow occurs and the dashed portion indicates the regime for which bidirectional flow occurs. b)-e) The unstable director, $\theta(z)$, (solid) and flow, $u(z)$, (dashed with arrows) modes for an activity distribution specified by the corresponding point in a).
    }
    \label{fig:Gradient}
\end{figure*}

From the equivalence to Schr\"{o}dinger's equation discussed in the previous section, we can anticipate the form of both the director and the flow without explicit calculation and in so doing provide various general statements. The director angle will exhibit sinusoidal behaviour in extensile regions and exponential decay in contractile ones, hence we expect the lowest-order director mode to have no internal nodes and so can take $\theta$ to be everywhere positive. If the system is purely extensile then from \eqref{eq:FlowDirector} the flow will always be positive and increasing up the film. If there is contractile material at the substrate then the flow will initially be negative, and extensile material closer to the free surface will result in bidirectional flow. If contractility only occurs higher in the film it will still reduce the flow speed, but bidirectionality will be contingent on whether it can overcome the flow induced by the preceding extensile region. This intuitive reasoning can be applied even in cases where there is no average activity gradient, such as when considering random variations in activity between agents. 

However, the appeal of a constant activity gradient is that it permits us to pass beyond qualitative description to a full solution in terms of Airy functions \cite{NIST:DLMF}. We write the activity as
\begin{equation}
    \zeta(z)=\bar{\zeta}+\frac{\Delta\zeta}{d}\left(z-\frac{d}{2}\right),
    \label{eq:LinearActivity}
\end{equation}
where $\bar{\zeta}$ is the average activity in the active film and $\Delta\zeta$ is the activity difference between the substrate and free surface, with positive or negative $\Delta\zeta$ meaning activity is increasing or decreasing with height through the film. Note that for the activity to be of the same sign everywhere in the film, either purely extensile or purely contractile, requires $|\Delta\zeta|<2|\bar{\zeta}|$. For example, in Figure \ref{fig:Gradient}a) the region of purely extensile activity is bounded by straight lines from the origin with gradient $\pm 2$, which pass through points $\text{c})$ and $\text{d})$.

Equation \eqref{eq:SchrodingerDirector} can then be solved to give the director angle
\begin{equation}
    \begin{split}
        \theta(z)&=C_1 \text{Ai}\left[-\left(\uppi\frac{\zeta_{\text{c}}}{\Delta\zeta}\right)^{2/3}\frac{\zeta(z)}{\zeta_{\text{c}}}\right]\\
        &\qquad +C_2 \text{Bi}\left[-\left(\uppi\frac{\zeta_{\text{c}}}{\Delta\zeta}\right)^{2/3}\frac{\zeta(z)}{\zeta_{\text{c}}}\right],
    \end{split}
\end{equation}
where $\text{Ai}$ and $\text{Bi}$ are the Airy functions of the first and second kind respectively and it is to be understood that the argument of each function is real.
Simultaneously satisfying both director boundary conditions requires
\begin{equation}
    \begin{split}
        &\text{Ai}\left[-\left(\uppi\frac{\zeta_{\text{c}}}{\Delta\zeta}\right)^{2/3}\frac{\zeta(d)}{\zeta_{\text{c}}}\right]\text{Bi}\left[-\left(\uppi\frac{\zeta_{\text{c}}}{\Delta\zeta}\right)^{2/3}\frac{\zeta(0)}{\zeta_{\text{c}}}\right]=\\
        &\ \ \ \text{Ai}\left[-\left(\uppi\frac{\zeta_{\text{c}}}{\Delta\zeta}\right)^{2/3}\frac{\zeta(0)}{\zeta_{\text{c}}}\right]\text{Bi}\left[-\left(\uppi\frac{\zeta_{\text{c}}}{\Delta\zeta}\right)^{2/3}\frac{\zeta(d)}{\zeta_{\text{c}}}\right],
    \end{split}
\end{equation}
defining a curve in $(\bar{\zeta},\Delta\zeta)$ space along which the linear instability occurs, shown in Figure \ref{fig:Gradient}a). Here the activity only appears when normalised by the critical value for a homogeneous system, $\zeta_{\text{c}}$ in \eqref{eq:homogeneousInstability}, making the instability curve shown in Figure \ref{fig:Gradient}a) universal and meaning that the transition to the unstable mode shown in Figure \ref{fig:homogeneousInstability} occurs at the point $(1,0)$, marked in Figure \ref{fig:Gradient}a) by an open circle. As noted above, in Figure \ref{fig:Gradient}a), the region between points $\text{c})$ and $\text{d})$ corresponds to a purely extensile system, in which we can see that increasing heterogeneity, i.e.~increasingly positive or negative $\Delta\zeta$, decreases the critical value of the average activity, $\bar{\zeta}$, so that there is a decrease in the total activity required for a flow transition. Outside this region the system contains contractile regions, with $\zeta<0$ and, although $\bar{\zeta}$ continues to decrease along the instability curve, this should not be interpreted as a reduction in the critical total activity. The appropriate measure of critical total activity is the integral of $|\zeta|$ through the film, which we denote by $\zeta_{\text{T}}^*$, and this is indeed increased by the presence of contractile stresses, as expected. The dashed portion of the instability curve in Figure \ref{fig:Gradient}a) for $\Delta\zeta>0$ corresponds to contractility near the substrate and produces the bidirectional flow shown in Figure \ref{fig:Gradient}b). When $\Delta\zeta<0$, contractile stresses are located at the free surface and flows of the type shown in Figure \ref{fig:Gradient}e) result. These features are explained by our previous reasoning, with the additional statement that for linear activity the shape of the instability curve and the exponential decay of the director angle in contractile regions precludes bidirectional flow from occurring when the extensile region is at the substrate.

Since this is a linear stability analysis, the modes depicted in \ref{fig:Gradient}b)-e) have arbitrary amplitude, although the director and flow modes maintain a common amplitude ratio. 
We can see from Figure \ref{fig:Gradient}b) and e) that a negative activity gradient enhances the magnitude of the flow relative to that of the director. This can be explained as a consequence of the symmetry of the system, as follows. The mirror symmetry of the instability curve and the forms of the director for $(\bar{\zeta},\pm\Delta\zeta)$ are immediately understood from the fact that $\Delta\zeta\to-\Delta\zeta$ and $z\to d-z$ is a symmetry of \eqref{eq:SchrodingerDirector} with activity given by \eqref{eq:LinearActivity}. Using $u^{\pm}$ to denote the flows resulting from equal and opposite activity gradients, employing these symmetries in \eqref{eq:FlowDirector} implies that
\begin{equation}
    u^-(z)=u^+(d)-u^+(d-z).
    \label{eq:ActivityGradientFlowSymmetry}
\end{equation}
It follows from the preceding discussion that within the dashed region of Figure \ref{fig:Gradient}a) $u^+(z)$ will have a flow minimum, where the flow is negative, at some point $z=z^*$ and achieve a maximum flow at the free surface. From \eqref{eq:ActivityGradientFlowSymmetry} we see that upon reversing $\Delta\zeta$ this flow minimum is mapped to an internal flow maximum at $d-z^*$. This relationship can be seen in the flows of Figure \ref{fig:Gradient}b) and e).

\subsection{Layered active systems}
\label{subsec:Layered active systems}
We now turn our attention to spontaneous flow transitions that occur in systems composed of layers of two distinct active nematics, as is relevant for the coexistence of different bacterial species or cellular interactions \cite{meacock2021bacteria,balasubramaniam2021investigating}. We take the activity to be uniform within each layer and focus on the two simplest scenarios: a configuration with a single interface between a lower and upper layer; and a sandwich configuration in which one active species encloses a layer of another. Following the correspondence with Schr\"{o}dinger's equation that we established earlier, these situations mirror potential steps and potential wells or barriers respectively.

\subsubsection{Activity steps}
\begin{figure*}
    \centering
    \begin{tikzpicture}[scale=1.6,>=stealth]
        \node[anchor=south west,inner sep=0] at (0,0)
{\includegraphics[width=0.95\linewidth, trim = 0 0 0 0, clip, angle = 0, origin = c]{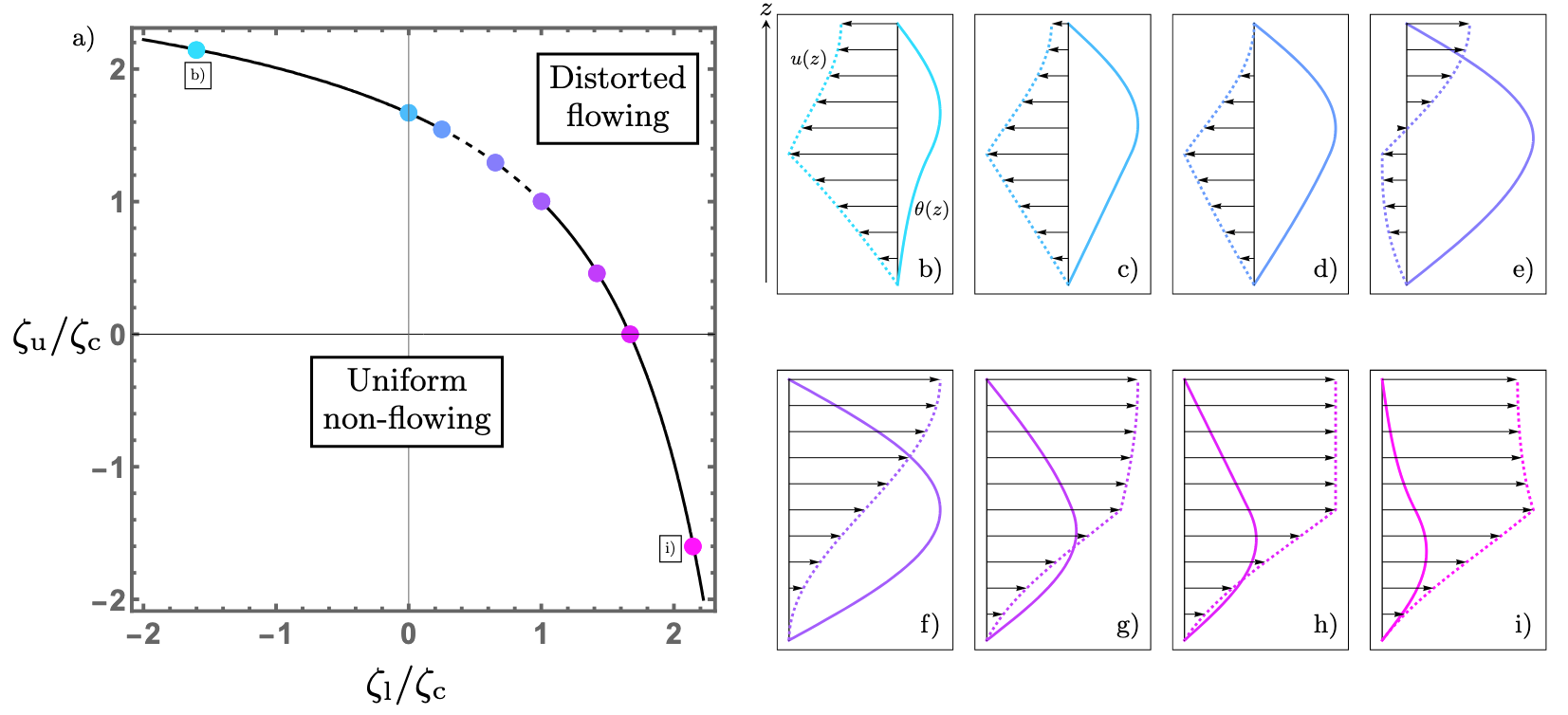}};
    \end{tikzpicture}
    \caption[The spontaneous flow transition of a two-component active nematic film.]{The spontaneous flow transition of a two-component active nematic film, composed of lower and upper layers with equal widths and activities $\zeta_{\text{l}}$ and $\zeta_{\text{u}}$ respectively. a) The line along which linear instability occurs. The solid portion indicates the regime for which unidirectional flow occurs and the dashed portion indicates the regime for which bidirectional flow occurs. b)-i) The unstable director, $\theta(z)$, (solid) and flow, $u(z)$, (dashed with arrows) modes for an activity distribution specified by the corresponding point in a). Panel f) shows the unstable modes for homogeneous activity, also illustrated in Figure \ref{fig:homogeneousInstability}.
    }
    \label{fig:Stack}
\end{figure*}

Beginning with the single activity step, it is elementary to solve \eqref{eq:SchrodingerDirector} to give the director as
\begin{equation}
    \theta(z)=\begin{cases}
        (-\text{i})^{\chi_{\text{l}}}A_{\text{l}}\sin\left(\uppi\sqrt{\dfrac{\zeta_{\text{l}}}{\zeta_{\text{c}}}}\dfrac{z}{d}\right) & 0\leq z \leq Z\\[1em]
        (-\text{i})^{\chi_{\text{u}}}A_{\text{u}}\sin\left(\uppi\sqrt{\dfrac{\zeta_{\text{u}}}{\zeta_{\text{c}}}}\left[\dfrac{z}{d}-1\right]\right) & Z \leq z \leq d
    \end{cases},
\end{equation}
where the subscripts indicate properties of the lower and upper regions, which have an interface at $z=Z$. The factors of $(-\text{i})^{\chi_{\text{i}}}$ are merely used to combine the possibilities of oscillatory or decaying behaviour in each region according to whether it is extensile ($\chi_{\text{i}}=0$) or contractile ($\chi_{\text{i}}=1$). As in solving Schr\"{o}dinger's equation for a potential step, we require continuity of the director and its derivative at the interface; which in the case of nematics  correspond to torque and force balances respectively. Satisfying this pair of conditions gives the following criterion for linear instability
\begin{equation}
    \sqrt{\frac{\zeta_{\text{l}}}{\zeta_{\text{u}}}}=\frac{\tan\left(\uppi\sqrt{\dfrac{\zeta_{\text{l}}}{\zeta_{\text{c}}}}\dfrac{Z}{d}\right)}{\tan\left(\uppi\sqrt{\dfrac{\zeta_{\text{u}}}{\zeta_{\text{c}}}}\left[\dfrac{Z}{d}-1\right]\right)}.
    \label{eq:StackedInstabCurve}
\end{equation}
The associated curve is shown in Figure \ref{fig:Stack}a) for the case of equal layer thicknesses, so that $Z=d/2$. It is now the point $(1,1)$ in Figure \ref{fig:Stack}a) that corresponds to a critically unstable homogeneous system with modes shown in Figure \ref{fig:homogeneousInstability}, and the expected symmetry under interchange of $\zeta_{\text{l}}$ and $\zeta_{\text{u}}$ is apparent. The total activity required for a flow transition is given by the sum of the activities in each layer, captured by the distance of the curve from the origin in a $L^1$-norm. As before, the critical activity is reduced by heterogeneities within an extensile system but increased once contractile stresses are present. It is minimal for an extensile-passive system.

The director solution is complemented by the following relation between the amplitudes 
\begin{equation}
    \frac{A_{\text{u}}}{A_{\text{l}}}=\frac{(-i)^{\chi_{\text{l}}-\chi_{\text{u}}}\sin\left(\uppi\sqrt{\dfrac{\zeta_{\text{l}}}{\zeta_{\text{c}}}}\dfrac{Z}{d}\right)}{\sin\left(\uppi\sqrt{\dfrac{\zeta_{\text{u}}}{\zeta_{\text{c}}}}\left[\dfrac{Z}{d}-1\right]\right)}.
\end{equation}
Similarly to before, the form of the director may be understood from the association between extensile and contractile materials and classically allowed and forbidden regions. The transformation $\zeta_{\text{l}}\leftrightarrow\zeta_{\text{u}}$ and $z\to d-z$ is a symmetry of the system and is reflected in Figure \ref{fig:Stack}b), c), h) and i).

The flow is no longer given merely by \eqref{eq:FlowDirector}, since steps in activity necessitate steps in flow gradients, and hence additional shear flows, in order to maintain stress continuity. We will briefly present the flow solution for an arbitrary number $n$ of active regions. Intergration of \eqref{eq:FlowStress} leads to the flow in each layer,
\begin{equation}
    \mu u_{\text{i}}=-\int_0^z\sigma_{xz}\text{d}z'+\alpha_{\text{i}}z+\beta_{\text{i}},
\end{equation}
where $\alpha_{\text{i}}$ and $\beta_{\text{i}}$ are the as-yet undetermined integration constants. For compactness in what follows we drop the component subscripts and denote $\sigma_{xz}$ by $\sigma$, using a subscript only to denote the region in which a field exists. At the interface between layers $\text{i}$ and $\text{i}+1$, located at $Z_{\text{i}}$, there is a jump in the elastic and active stress
\begin{equation}
    \Delta\sigma_{\text{i}}=\sigma_{\text{i}+1}(Z_{\text{i}})-\sigma_{\text{i}}(Z_{\text{i}}),
\end{equation}
with similar definitions for the jumps in the constants $\Delta\alpha_{\text{i}}$ and $\Delta\beta_{\text{i}}$. Stress continuity then requires
\begin{equation}
    \Delta\alpha_{\text{i}}=-\Delta\sigma,
\end{equation}
and flow continuity requires
\begin{equation}
    \Delta\beta_{\text{i}}=Z_{\text{i}}\Delta\sigma_{\text{i}}.
\end{equation}
With the addition of the no-slip condition at the substrate (so $\beta_1=0)$ and vanishing of viscous stress at the free surface (so $\alpha_\text{n}=0$) we determine the flow to be given by
\begin{equation}
    u_{\text{i}}=\frac{1}{\mu}\left[-\int_0^z\sigma\text{d}z'+z\sum_{j=i}^{n-1}\Delta\sigma_{\text{j}}+\sum_{j=1}^{i-1}Z_{\text{j}}\Delta\sigma_{\text{j}}\right].
    \label{eq:FlowJumpsGeneric}
\end{equation}

Restricting to the case of two layers and a single interface, we see that the fluid flow is given by
\begin{equation}
    u(z)=\begin{cases}
        \frac{1}{\bar{\mu}}\displaystyle\int_0^z\zeta\theta\text{d}z'-z\theta(Z)\Delta\zeta, & 0\leq z \leq Z\\[1em]
        \frac{1}{\bar{\mu}}\displaystyle\int_0^z\zeta\theta\text{d}z'-Z\theta(Z)\Delta\zeta, & Z\leq z \leq d
    \end{cases},
\end{equation}
where $\bar{\mu}=\mu+(\gamma(1-\nu)^2)/4$ and now $\Delta\zeta=\zeta_{\text{u}}-\zeta_{\text{l}}$. The flows for different values of $\zeta_{\text{u}}$ and $\zeta_{\text{l}}$ are shown as dashed lines with arrows in Figure \ref{fig:Stack}b)-i). In the upper layer the sign of the flow gradient is determined solely by the nature of the activity there. It is positive when the region is extensile, as seen in Figure \ref{fig:Stack}b)-g), zero when it is passive, resulting in the plug flow shown in Figure \ref{fig:Stack}h), and negative when it is contractile, illustrated in Figure \ref{fig:Stack}i). In all cases the magnitude of the flow gradient decreases towards the free surface in the upper layer due to the vanishing of the director angle. This is particularly pronounced in the contractile case due to the exponential suppression of the director distortion.

The situation is different in the lower layer, where the activity jump induces a shear flow and the flow gradient is proportional to $\zeta_{\text{l}}\theta(z)-\Delta\zeta\theta(Z)$. When $\zeta_{\text{l}}>0$ and $\Delta\zeta>0$ and sufficiently small, that is a fully extensile system with a more active layer on top, bidirectional flow occurs, indicated by the dashed line in Figure \ref{fig:Stack}a). When $\Delta\zeta$ is small the flow reversal will happen near the substrate and as $\Delta\zeta$ is increased the point of flow reversal moves upwards through the film, as illustrated in Figure \ref{fig:Stack}e). The bidirectional flow regime ceases when the flow at the free surface vanishes, shown in Figure \ref{fig:Stack}d), that is when
\begin{equation}
    \frac{Z}{d}\theta(Z)\Delta\zeta=\frac{1}{d}\int_0^d\zeta\theta\text{d}z.
    \label{eq:StackedBi-directFlow}
\end{equation}
This interface-driven bidirectional flow is in contrast with the case of smooth activity variations, where both extensile and contractile regions are necessary for bidirectionality.

It should be noted that although the flow for the unstable mode along both solid parts of the line of instability in Figure \ref{fig:Stack}a) is unidirectional, it is in opposing senses. In the solid section for high values of $\zeta_{\text{u}}$ the flow has the opposite sign as the director, seen in Figure \ref{fig:Stack}b)-d), while in the solid section for low values of $\zeta_{\text{u}}$ it has the same sign, shown in Figure \ref{fig:Stack}f)-i). This means that interchanging the activities of the two regions can reverse the direction of flow. It also means that within the dashed portion of the curve in Figure \ref{fig:Stack}a) there is a point with no net flux along the film.

In Figure \ref{fig:Stack} we show the form of the transition only for equal width layers, but changing the height of the interface, $Z$, leads to no new phenomena and the essential reasoning from above remains the same. When $Z\ne d/2$ the instability curve in Figure \ref{fig:Stack}a) will not be symmetric, but rather stretched in the direction of the activity of the narrower layer according to \eqref{eq:StackedInstabCurve}. Similarly, the regime of bidirectional flow, while always present, will grow or shrink according to \eqref{eq:StackedBi-directFlow}.

\subsubsection{Activity wells and barriers}
\begin{figure*}
    \centering
    \begin{tikzpicture}[scale=1.6,>=stealth]
        \node[anchor=south west,inner sep=0] at (0,0)
{\includegraphics[width=0.95\linewidth, trim = 0 0 0 0, clip, angle = 0, origin = c]{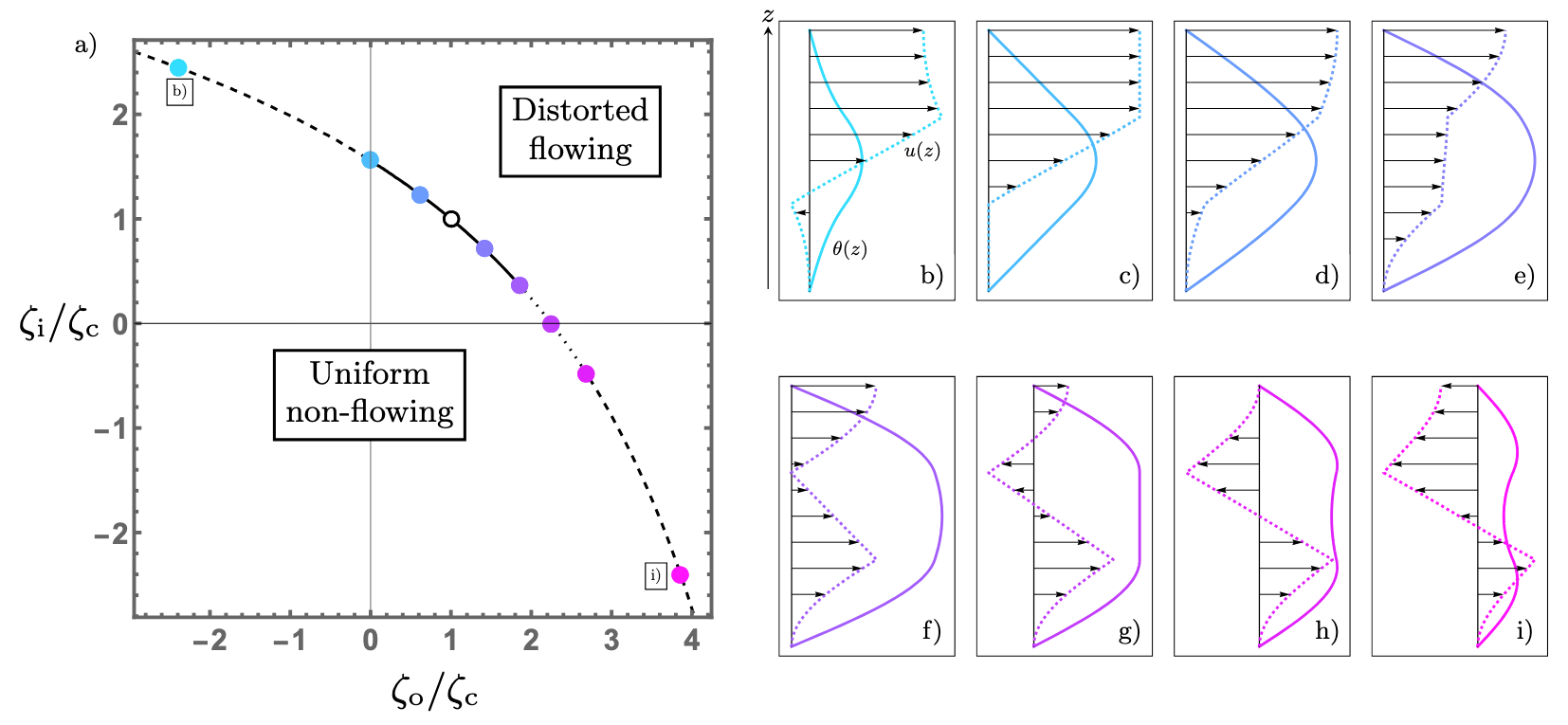}};
    \end{tikzpicture}
    \caption[The spontaneous flow transition of a two-component active nematic film.]{The spontaneous flow transition of a two-component active nematic film, composed of two outer regions of activity $\zeta_{\text{o}}$ enclosing an inner region with activity $\zeta_{\text{i}}$, with all regions having equal widths. The open circle corresponds to the critical activity in a homogeneous system. a) The line along which linear instability occurs. The solid, dashed and dotted portions indicate the regimes for which unidirectional, bidirectional and three-fold flow occurs respectively. b)-i) The unstable director, $\theta(z)$, (solid) and flow, $u(z)$, (dashed with arrows) modes for an activity distribution specified by the corresponding point in a).
    }
    \label{fig:Sandwich}
\end{figure*}

The analysis proceeds similarly for the activity well and barrier system, where there is a sandwich configuration in which one active species encloses a layer of another. We will consider an even activity profile, with outer layers of activity $\zeta_{\text{o}}$ for $0<z<Z_1$ and $d-Z_1<z<d$ and an inner layer of activity $\zeta_{\text{i}}$  for $Z_1<z<d-Z_1$. The assumption of symmetry simplifies the analysis without sacrificing any important phenomena. The interfaces between layers therefore occur at $Z_1$ and $Z_2=d-Z_1$ and the jumps in activity are $\Delta\zeta_1=\zeta_{\text{i}}-\zeta_{\text{o}}\doteq\Delta\zeta$ and $\Delta\zeta_2=\zeta_{\text{o}}-\zeta_{\text{i}}=-\Delta\zeta$.  The director is given by
\begin{equation}
    \theta(z)=\begin{cases}
        (-\text{i})^{\chi_{\text{o}}}A_{\text{o}}\sin\left(\uppi\sqrt{\dfrac{\zeta_{\text{o}}}{\zeta_{\text{c}}}}\dfrac{z}{d}\right) & 0 \leq z \leq Z_1\\[1em]
        A_{\text{i}}\cos\left(\uppi\sqrt{\dfrac{\zeta_{\text{i}}}{\zeta_{\text{c}}}}\left[\dfrac{z}{d}-\dfrac{1}{2}\right]\right) & Z_1 \leq z \leq d-Z_1\\[1em]
        (-\text{i})^{\chi_{\text{o}}}A_{\text{o}}\sin\left(\uppi\sqrt{\dfrac{\zeta_{\text{o}}}{\zeta_{\text{c}}}}\left[\dfrac{z}{d}-1\right]\right) \hspace*{-0.15cm}& d-Z_1 \leq z \leq d
    \end{cases},
\end{equation}
where the subscripts denote properties of the outer and inner regions. As before, requiring continuity of $\theta$ and its derivative at $z=Z_1$ and $z=d-Z_1$ determines a curve along which the linear instability occurs together with a coupling of the amplitudes, specified by
\begin{align}
    \sqrt{\frac{\zeta_{\text{o}}}{\zeta_{\text{i}}}}&=-\tan\left(\uppi\sqrt{\frac{\zeta_{\text{o}}}{\zeta_{\text{c}}}}\frac{Z_1}{d}\right)\tan\left(\uppi\sqrt{\frac{\zeta_{\text{i}}}{\zeta_{\text{c}}}}\left[\frac{Z_1}{d}-\frac{1}{2}\right]\right)
    \label{eq:InstabilityConditionSandwich}
\end{align}
and
\begin{align}
    \dfrac{A_{\text{i}}}{A_{\text{o}}}&=\frac{(-\text{i})^{\chi_{\text{o}}}\sin\left(\uppi\sqrt{\dfrac{\zeta_{\text{o}}}{\zeta_{\text{c}}}}\dfrac{Z_1}{d}\right)}{\cos\left(\uppi\sqrt{\dfrac{\zeta_{\text{i}}}{\zeta_{\text{c}}}}\left[\dfrac{Z_1}{d}-1\right]\right)}
\end{align}
respectively. Figure \ref{fig:Sandwich}a) shows the curve along which the linear instability occurs for the case of equal layer thicknesses, $Z_1=d/3$. The principal new feature compared to before is that when the inner layer is contractile, so that $\zeta_{\text{i}}<0$, the unstable director mode exhibits a pair of maxima, as illustrated in Figure \ref{fig:Sandwich}h) and i). We shall return to this point at the end of this section.

Since the interfaces only induce shear flow in the layers below them, we can build up an understanding of multi-layered systems iteratively. For this three-layered configuration the flow in the upper two layers follows the same principles as laid out in our previous discussion of the two-layer, single-activity step system, with two caveats. The first is that the flow at the bottom of the inner layer does not vanish, but rather has a value determined by the stresses in the lowest layer. This velocity enters as an additive constant and so only modifies our previous description by a Galilean boost. Secondly, the director is now also non-zero at the bottom of the inner layer, altering the competition between the local active stress and induced shear contributions to the flow contained in \eqref{eq:FlowJumpsGeneric}. The additional bottom layer has a flow that, following \eqref{eq:FlowJumpsGeneric}, involves a shear component set by both activity jumps. Owing to the symmetry of the activity profile we present here, this shear contribution vanishes and the flow in each layer may be written as
\begin{widetext}
    \begin{equation}
    u(z)=\begin{cases}
        \dfrac{1}{\bar{\mu}}\displaystyle\int_0^z\zeta\theta\text{d}z' & 0 \leq z \leq Z_1\\[1em]
        \dfrac{1}{\bar{\mu}}\left[\displaystyle\int_0^z\zeta\theta\text{d}z'+(z-Z_1)\theta(Z_1)\Delta\zeta\right] & Z_1 \leq z \leq d-Z_1\\[1em]
        \dfrac{1}{\bar{\mu}}\left[\displaystyle\int_0^z\zeta\theta\text{d}z'+(d-2Z_1)\theta(Z_1)\Delta\zeta\right] & d-Z_1 \leq z \leq d
    \end{cases}.
    \label{eq:FlowSandwich}
\end{equation}
\end{widetext}
Note that the flow gradient in both outer layers is set purely by the form of activity there, with an interface-induced shear flow only featuring in the inner layer. This shear component in the inner layer is a key driver of distinct flow morphologies.

We begin with an overview of these possible flow morphologies at instability. There is a single unidirectional flow regime, two bidirectional flow regimes and one three-fold flow regime, indicated in Figure \ref{fig:Sandwich}a)  repectively via solid, dashed and dotted sections of the instability curve. The net flow is positive, in the same sense as the director, to the left of the three-fold region and negative to the right, and is zero at a point within the three-fold region.

When two contractile layers enclose an extensile layer ($\zeta_{\text{i}}>0,\,\zeta_{\text{o}}<0$), forming an activity well, the flow gradient will be negative at the substrate, with a large positive gradient in the inner layer, producing bidirectional flow, as seen in the leftmost dashed section in Figure \ref{fig:Sandwich}a) and illustrated in Figure \ref{fig:Sandwich}b). If all regions are extensile  and the inner layer is more active ($\zeta_{\text{i}}>\zeta_{\text{o}}>0$) then all contributions to the flow gradients are positive and the flow is unidirectional and increasing in magnitude up the film from substrate to free surface, as illustrated in Figure \ref{fig:Sandwich}d). However, if the activity is larger in the outer layers ($\zeta_{\text{o}}>\zeta_{\text{i}}$)  the shear term in \eqref{eq:FlowSandwich} becomes negative and if this is large enough then three-fold flow results, as seen in Figure \ref{fig:Sandwich}g). The transitions into and out of this regime are marked by the flow vanishing at the bottom of the top layer or at the free surface. These two scenarios are shown in Figure \ref{fig:Sandwich}f) and h). Thus three-fold flow arises when an extensile material surrounds a weakly active one, which may be extensile, passive or contractile. This may be thought of as a low activity-barrier. As the height of the barrier increases, that is, the inner material becomes more contractile, the increasing negativity of the shear flow in that layer results in another regime of bidirectional flow, illustrated in Figure \ref{fig:Sandwich}i).

For this three-layer system we can again consider the role of active heterogeneities in modifying the critical total activity required for a flow transition. This can not be inferred from Figure \ref{fig:Sandwich}a) simply from an $L^1$-distance, since the outer regions constitute two thirds of the material and so $\zeta_{\text{o}}$ must be weighted by a factor of two. Nonetheless, the situation is familiar as far as contractile stresses are concerned, their presence impedes the instability and increases the critical total activity. Within the purely extensile region we now find that the total activity threshold is only reduced compared to the homogeneous case by making the inner layer more extensile and is minimised for an extensile region enclosed by two passive ones.

\begin{figure}
    \centering
    \begin{tikzpicture}[scale=1.6,>=stealth]
        \node[anchor=south west,inner sep=0] at (0,0)
{\includegraphics[width=0.95\linewidth, trim = 0 0 0 0, clip, angle = 0, origin = c]{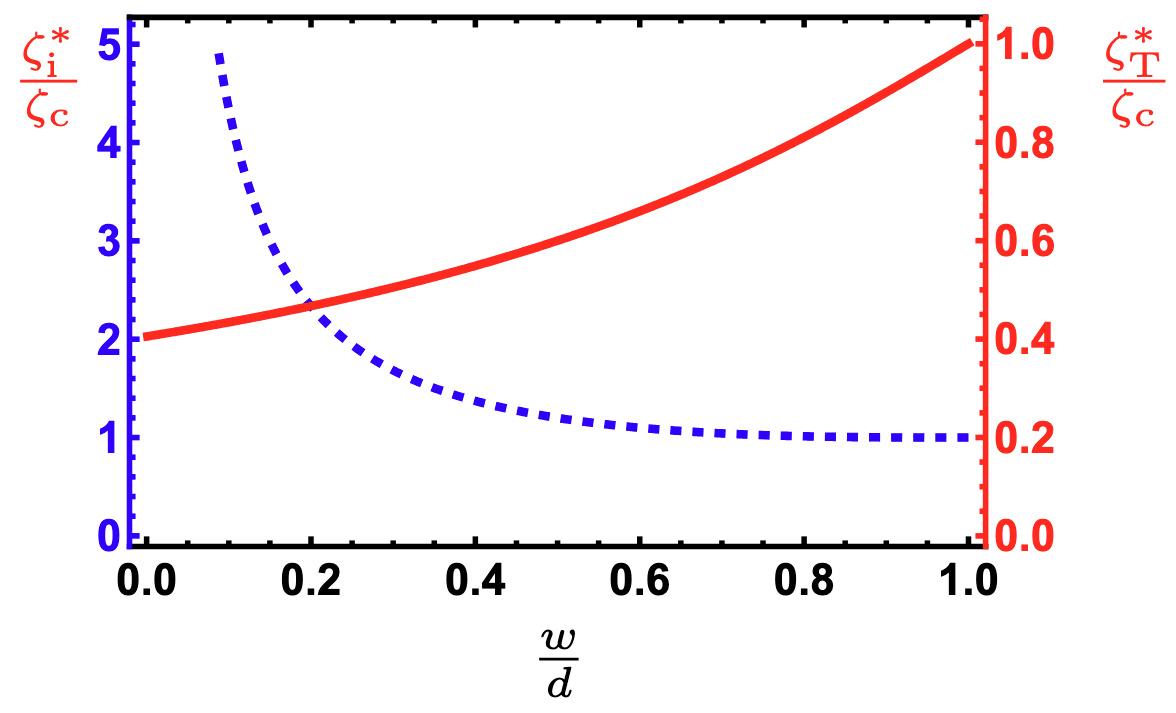}};
    \end{tikzpicture}
    \caption[The critical activity in an extensile-passive system.]{The critical activity for an inner extensile layer of width $w$ symmetrically enclosed by two passive layers of equal thickness, for a total film height of $d$. The critical activity value within the extensile region, $\zeta_{\text{i}}^*$ (blue, dashed), and the total threshold activity in the film, $\zeta_{\text{T}}^*$ (red, solid) are shown, normalised by their values for a homogeneous system, which corresponds to $\frac{w}{d}=1$. 
    }
    \label{fig:NetActivityReduction}
\end{figure}

We can investigate the importance of the inner layer in determining the activity threshold of this system by varying the width, $w$, of the inner extensile layer, with $w=d-2Z_1$. By taking the $\zeta_{\text{o}}\to0$ limit of \eqref{eq:InstabilityConditionSandwich}, we find the instability condition to be $\frac{d-w}{2d}\tan\left(\uppi\sqrt{\frac{\zeta_{\text{i}}^*}{\zeta_{\text{c}}}}\frac{w}{2d}\right)=\frac{1}{\uppi}\sqrt{\frac{\zeta_{\text{c}}}{\zeta_{\text{i}}^*}}$, where we use $\zeta_{\text{i}}^*$ to denote the critical activity in the inner layer. Naturally, this increases as the active layer shrinks, shown by the blue dashed curve in Figure \ref{fig:NetActivityReduction}. However, this narrowing also acts to reduce the total activity in the system and from the red solid curve in Figure \ref{fig:NetActivityReduction} we can see that this effect dominates. Although decreasing the size of the extensile layer increases the threshold activity value $\zeta_{\text{i}}^*$ in that region, it decreases the total activity required for criticality, $\zeta_{\text{T}}^*$, in this case given by $w\zeta_{\text{i}}^*$. This means that a redistribution of active agents within a sub- or super-critical system might induce or supress a spontaneous flow transition. To determine the strength of this effect we expand the above instability criterion to linear order in $w$ and find $\zeta_{\text{T}}^*$ in an infinitesimally narrow active layer to be $\frac{4\eta}{d}$. Comparing this with the homogeneous system value of $\frac{\uppi^2\eta}{d}$, given in  \eqref{eq:homogeneousInstability}, we see that focusing active agents to the centre of the film can reduce the total activity threshold by a factor of $\frac{4}{\uppi^2}$, or about $40\%$.

A previous connection between active matter and Schr\"{o}dinger's equation has been made in the description of inertial active particles \cite{te2023microscopic}, and the reduction of the director angle in the inner contractile region, seen in Figure \ref{fig:Sandwich}h) and i) mirrors the drop in particle density seen in that context as an `active tunnelling' effect. However, a notable difference here is that unfavourable regions of activity do not only suppress the director, but also drastically change the nature of the induced flow.

A key feature that runs through the results of this paper is that the nature of the flow transition is highly dependent not only on the types of activity present, but also on where there are located. This is clearly demonstrated by the two extensile-passive configurations in Figure \ref{fig:Sandwich}, namely the point c) where $\zeta_{\text{i}}>0,\,\zeta_{\text{o}}=0$ and the point g) where $\zeta_{\text{i}}=0,\,\zeta_{\text{o}}>0$. We see from Figure \ref{fig:Sandwich}c), when passive layers enclose an extensile layer, the flow is unidirectional, with plug flow in the outer regions. However, exchanging the roles of passive and extensile layers changes the system considerably and produces the three-fold flow shown in Figure \ref{fig:Sandwich}g).

\section{Conclusion}
We have investigated spontaneous flow transitions in active nematic films with heterogeneous activity.  Our results show that such heterogeneities can substantially alter the nature of the induced flow, including reversal or diminishing of the net fluid flux  and flows that change direction multiple times through the film due to shear induced by activity interfaces. These are both of direct biological relevance. The fluid flux determines the spread of nutrients along the film \cite{hall2004bacterial,thampi2022channel}, while shear flows impact the orintation and position of rheotactic microswimmers \cite{miki2013rheotaxis} and can cause the deformation or detachment of biofilms \cite{stoodley2002biofilm}. Throughout, we observe that these effects depend strongly not just on the types of activity present in the film, but on where they are located. Furthermore, the total activity in the film necessary to induce spontaneous flow can be decreased or increased by redistributing the active agents. Interestingly, in a channel rather than film geometry, both the instability-hindering role of contractile activity and the formation of shear flows due to active heterogeneity have been seen in simulations \cite{marenduzzo2010hydrodynamics}. Many of the predictions we have presented should be experimentally testable, most readily those concerning extensile-passive systems. All these results occurred against the backdrop of a correspondence between the condition for an unstable director mode and the time-independent Schr\"{o}dinger equation. While this is not the first work to connect active and quantum matter \cite{te2023microscopic,zheng2023quantum}, it opens up another avenue for the exchange of ideas between the two fields.

One might ask how the results we have presented here would alter if there was weak anchoring at the free surface. While we postpone a full investigation of this topic, we can make the following statements. Weakening the anchoring will monotonically lower the critical activity required for spontaneous flow, analogous to its role in the Freedericksz transition \cite{rapini1969distorsion}. It is known that any generic weak anchoring can always be treated by extending the system by a surface `extrapolation length' to a virtual boundary on which infinite anchoring is imposed \cite{de1993physics}. This extrapolation length can typically be viewed as inducing a phase shift to the director, with the remaining analysis then proceeding as presented here. The case of infinite anchoring therefore provides an overarching structure through which to understand all other anchoring conditions, where the symmetries that are manifest in this paper would appear in a modified fashion. This gives our results for infinite anchoring a special status, beyond simply corresponding to a particular choice of anchoring conditions.

In this paper we have not incorporated deformations of any interfaces, focusing on the effects due to distinct activities. This reduced description is appropriate if the variation in activity is imprinted, for example by modulating light intensity \cite{zhang2021spatiotemporal}. A natural application of our results is therefore to active nematic microfluidics, using activity patterning to control the existence and structure of a spontaneous flow, and we intend to present results for a channel geometry in a subsequent publication. However, we still expect our results to be of relevance in the biological context presented here, as has proven to be true in the homogeneous case \cite{voituriez2005spontaneous}. Nonetheless,  it will be interesting to consider a more sophisticated treatment of the interfaces, which could include active anchoring \cite{blow2017motility,ruske2022activity} and a coupling of the hydrodynamic instability to perturbations of the free surface \cite{sankararaman2009instabilities}.

A natural extension of our work would be to consider the effect of heterogeneous activity on the spontaneous flow transitions of three-dimensional active nematics. These have attracted considerable recent interest \cite{chandrakar2020confinement,singh2023spontaneous}, distinguished as they are from two-dimensional ones by the manner of the transition between turbulent and coherent flows \cite{wu2017transition,chandragiri2020flow,varghese2020confinement}, the ability to form chiral structures \cite{keogh2022helical,pratley2024three} and the existence of novel instabilities \cite{strubing2020wrinkling}.

The varied form of flow instabilities that we have presented hints at a richness in the non-linear effects that might lie beyond. Even in homogeneous systems, such non-linearities can lead to complex behaviour, including instabilities to higher dimensional distortions and topological defects. In regions of heterogeneous activity, topological defects   experience polarising torques \cite{ruske2022activity}, so that their spatio-temporal dynamics can be controlled through activity patterning \cite{zhang2021spatiotemporal,shankar2024design}, and in heterogeneous extensile systems the familiar defect-mediated transition to active turbulence \cite{alert2022active} via a Ceilidh dance state \cite{shendruk2017dancing} is supplemented by a regime of phase separation \cite{assante2023active}. However, little has been done in the directions of extensile-contractile mixtures or three-dimensional systems and so a computational challenge remains in determining the full, post transition, dynamics of heterogeneous active nematics.

\section*{Data availability}
All methods to produce the data  of the paper are available within the paper itself.

\bibliography{SchrodingerBiofilm.bib}

\section*{Acknowledgements}
For the purpose of open access, the authors have applied a Creative Commons Attribution (CC-BY) licence to any Author Accepted Manuscript version arising from this submission. The authors would like to acknowledge the financial support of the Medical Research Council [grant number G0902331] and the Engineering and Physical Sciences Research Council [grant number EP/T012501/2]. 

\section*{Author contributions}
A.H. and N.M. conceived of the research and methodology, A.H. carried out the calculations and analysis, A.H. and N.M. discussed the results and their interpretation, A.H. produced an initial version of the manuscript and A.H. and N.M. contributed to the editing and final version.

\section*{Competing interests}
The authors declare no competing interests.

\end{document}